\documentclass[aip,jcp,11pt,preprint,floatfix,showkeys]{revtex4-1}

\usepackage[T1]{fontenc}
\usepackage[utf8]{inputenc}
\usepackage{graphicx}
\usepackage{amsmath}
\usepackage[implicit=false,colorlinks=true,urlcolor=black,linkcolor=black,bookmarks=false,pdfpagelabels=false]{hyperref}

\begin{document}

\title{Insights into the Deactivation of 5-Bromouracil after UV Excitation}

\author{Francesca Peccati}
\thanks{These authors contributed equally to this work.}
\affiliation{Departament de Qu\'{\i}mica, Universitat Aut\`onoma de Barcelona, 08193 Bellaterra, Spain}

\author{Sebastian Mai}
\thanks{These authors contributed equally to this work.}
\affiliation{Institute of Theoretical Chemistry, Faculty of Chemistry, University of Vienna, W\"ahringer Stra\ss{}e 17, 1090 Vienna, Austria}

\author{Leticia Gonz\'alez}
\email{leticia.gonzalez@univie.ac.at}
\affiliation{Institute of Theoretical Chemistry, Faculty of Chemistry, University of Vienna, W\"ahringer Stra\ss{}e 17, 1090 Vienna, Austria}



\date{\today}
\begin{abstract}

5-Bromouracil is a nucleobase analogue that can replace thymine in DNA strands and acts as a strong radiosensitizer, with potential applications in molecular biology and cancer therapy.
Here, the deactivation of 5-bromouracil after UV irradiation is investigated in the singlet and triplet manifold by accurate quantum chemistry calculations and nonadiabatic dynamics simulations.
It is found that after irradiation to the bright $\pi\pi^*$ state, three main relaxation pathways are in principle possible: relaxation back to the ground state, intersystem crossing, and C-Br photodissociation.
Based on accurate MS-CASPT2 optimizations, we propose that ground state relaxation should be the predominant deactivation pathway in gas phase.
We then employ different electronic structure methods to assess their suitability to carry out excited-state dynamics simulations.
MRCIS was used in surface hopping simulations to compute the ultrafast intersystem crossing dynamics, which mostly involves the $^1n_\text{O}\pi^*$ and $^3\pi\pi^*$ states.

\end{abstract}

\maketitle


\section{Introduction}

\begin{figure}
  \centering
  \includegraphics[scale=1]{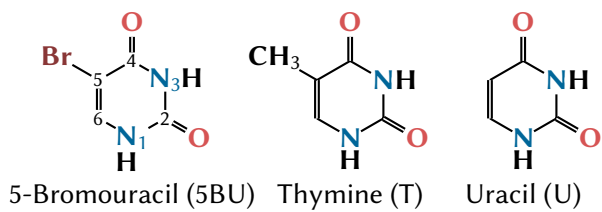}
  \caption{Chemical structures of 5-bromouracil (with ring atom numbering), thymine, and uracil.}
  \label{fig:scheme1}
\end{figure}

5-Bromouracil (5BU) can replace thymine (T) in DNA,\cite{Dunn1954N} causing strong mutagenic effects that have been explained by base mispairings due to 5BU's tautomeric equilibrium.\cite{Benzer1958PNAS,Orozco1998JPCB}
The substitution of T by 5BU is also accompanied by an increased UV light sensitivity of the affected DNA,\cite{Djordjevic1960JEM} where the presence of the 5BU can cause single strand breaks, alkali-labile bonds, double strand breaks, and DNA-protein crosslinking.\cite{Hutchinson1973QRB, Sugiyama1990JACS,Dietz1987JACS}
The occurrence of these lesions originates from the ability of 5BU to cleave the C-Br bond, leading to the formation of an uracilyl radical.\cite{Campbell1974ZNB,Rothman1967PP}
This radical in turn can abstract a hydrogen atom from a suitable donor (e.g., an adjacent sugar), thereby forming uracil (U); see Figure~\ref{fig:scheme1} for the chemical structures of 5BU, T, and U.
Hence, the uracilyl radical is responsible for DNA damage and can be exploited for a variety of processes, such as DNA crosslinking,\cite{Dietz1987JACS} adenine elimination,\cite{Sugiyama1990JACS} or generation of reactive oxygen species for photodynamical therapy.

Two different mechanisms have been proposed for the dissociation of the C-Br bond of 5BU.\cite{Swanson1981JACS,Dietz1987JACS,Storoniak2011JPCB}
The first route is initiated by an excited-state electron transfer from an adjacent nucleobase to 5BU, resulting in the formation of the 5BU$^{\cdot -}$ radical anion, which then undergoes C-Br cleavage, liberating a bromide ion and leaving behind the highly reactive uracilyl radical.\cite{Chen2000JACS}
In the second one, 5BU can also homolytically cleave the C-Br bond without any electron transfer involved.\cite{Campbell1974ZNB,Rothman1967PP}
These two routes differ in their selectivity for certain types of damages.\cite{Dietz1987JACS}
Moreover, it has been shown that the damage induced by 5BU depends on the DNA conformation,\cite{Dextraze2007B} which could be utilized in tumor radiosensitization applications.\cite{Prados1999IJROBP}

Given the potential applicability of 5BU, which has been even used in a clinical trial,\cite{Prados1999IJROBP} a number of papers focused on the primary photochemical properties of 5BU.
However, the molecular details of the deactivation of 5BU are still poorly understood.
Based on experimental evidence, it has been suggested that C-Br homolysis occurs in the excited singlet state (starting from $^1\pi\pi^*$).\cite{Swanson1981JACS}
Intersystem crossing (ISC) to the triplet manifold has been also reported,\cite{Swanson1981JACS,Dietz1987JACS} but in these experiments performed in 2-propanol, dissociation takes place involving electron transfer from the solvent.
In this case, the associated quantum yields were reported to be 1.3\% for homolysis from $^1\pi\pi^*$, 3\% for ISC, and 0.6\% for ISC followed by dissociation.\cite{Swanson1981JACS,Dietz1987JACS}
Excitation at lower energy apparently increases the ISC yield to 6\%.\cite{Dietz1987JACS}

5BU has also been studied with femtosecond time-resolved pump-probe experiments in water,\cite{Lu2004JPCB,Wang2006JCPa} where it has been found that after excitation, most of the signal decays with a 400~fs time constant, and the long-lived residual decays with a nanosecond time constant.
Since the work reported in these studies\cite{Lu2004JPCB,Wang2006JCPa} was focused on the dynamics of solvated electrons and the formation of 5BU$^{\cdot -}$, no further interpretation was given regarding the time scales intrinsic to 5BU.

Mechanistically, static quantum chemical calculations on isolated 5BU have provided support for two pathways operating after 5BU is irradiated by UV light---one pathway is responsible for bromine elimination and the second leads back to the reactant through a conical intersection.\cite{Kobylecka2009JPCA}
According to that study, there are two internal coordinates that are critical for the description of the deactivation processes of 5BU, the C$_5$-Br bond length and the out-of-plane angle of the bromine atom.
The authors then proposed that reactivity is controlled by an extended $S_0/\pi\pi^*$ near-degeneracy seam along a combination of these two internal coordinates.
This seam can be accessed after small barriers are overcome.
Depending on the particular point on the seam at where the system decays to the ground state, either dissociation or reactant recovery takes place.

In this paper, we study the gas-phase deactivation pathways of 5BU after UV irradiation using potential energy surface explorations and additionally using the \textsc{Sharc} (Surface Hopping with ARbitrary Couplings) dynamics method,\cite{Richter2011JCTC,Mai2015IJQC} in order to get additional insights, paying special attention to the role of triplet states.
Previous \textsc{Sharc} simulations have shown that triplet states can be populated in ultrafast short time scales in a number of nucleobases\cite{Mai2013C, Richter2014PCCP} and nucleobases analogues, such as those resulting from the substitution of the oxygen atom by the heavier sulfur atom.\cite{Mai2016JPCL,Mai2016NC,Mai2016CP}
Here, the presence of the heavy bromine atom might provide strong spin-orbit couplings making the intersystem crossing (ISC) to the triplet states a viable process, as proposed experimentally.\cite{Swanson1981JACS,Dietz1987JACS}


\section{Computational Methods}

Two different strategies were employed here to study the deactivation pathways of 5BU, static quantum chemical calculations (single-point calculations, optimizations and linear interpolation scans) and nonadiabatic dynamics simulations.
The methods employed in each case are detailed in the following.

\subsection{Quantum chemistry calculations}

\begin{figure}
  \centering
  \includegraphics[scale=1]{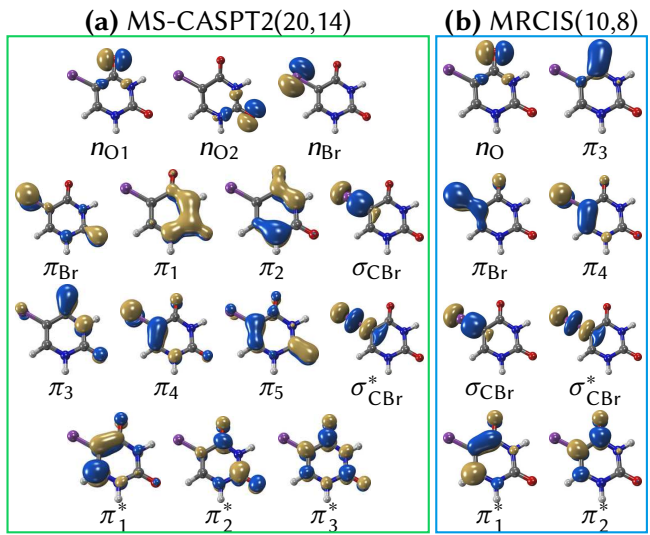}
  \caption{
  (a) Active space orbitals of the MS-CASPT2(20,14)/ANO-RCC-VDZP calculation.
  (b) Reference space orbitals of the MRCIS(10,8)/cc-pVDZ-DK calculations.
  }
  \label{fig:orbs}
\end{figure}

Vertical excitation calculations were performed, using the ground state geometry reported in Ref.~\onlinecite{Kobylecka2009JPCA}, which was optimized at the CASSCF(12,10)/6-311G* level of theory.
First, the vertical excitation calculations were performed with \textsc{Molcas} 8.0\cite{Aquilante2015JCC} at the MS-CASPT2(20,14) level of theory.
We employed the ANO-RCC-VDZP\cite{Roos2004JPCA} basis set, an IPEA shift\cite{Ghigo2004CPL} of zero\cite{Zobel2016CS} and an imaginary level shift\cite{Finley1998CPL} of 0.2~a.u. to exclude intruder states.
The employed active space orbitals (9 $\pi/\pi^*$ orbitals, 3 lone pair orbitals, and the $\sigma/\sigma^*$ pair of the C-Br bond) are shown in Figure~\ref{fig:orbs}a.
Either 9 singlet states or 8 triplet states were included in the calculations, where this relatively large number of states was necessary to stabilize the active space.

In order to find a cheaper but still accurate alternative to MS-CASPT2(20,14), vertical excitation calculations were then performed with additional methods, using the same geometry.
Here, we employed SA-CASSCF(14,11)/cc-pVDZ-DK\cite{Jong2001JCP} (using \textsc{Molcas} 8.0\cite{Aquilante2015JCC}), ADC(2)/def2-TZVP (using \textsc{Turbomole} 7.0\cite{TURBOMOLE70}), TD-BP86/TZP, TD-B3LYP/TZP (both using ADF\cite{ADF2014}), and MRCIS/cc-pVDZ-DK\cite{Jong2001JCP} (using \textsc{Columbus} 7\cite{Lischka2011WCMS,Lischka2012a}).

The three main relaxation pathways possible in 5BU---ground state relaxation, ISC, and photodissociation---were investigated by means of optimizations of excited-state minima and crossing points.
To this aim, we employed the external optimizer of ORCA,\cite{Neese2012WCMS} which was fed the appropriate gradients according to Refs.~\onlinecite{Levine2008JPCB, Bearpark1994CPL}.
Due to the high cost of MS-CASPT2(20,14) calculations, these optimizations employed gradients from ADC(2) or numerical gradients from MS-CASPT2(12,9).
Based on the optimized geometries, several linear-interpolation-in-internal-coordinates (LIIC) scans were performed at the MS-CASPT2(20,14) and ADC(2) levels of theory.

\subsection{Nonadiabatic dynamics simulations}

Dynamics simulations were carried out with a local development version of the \textsc{Sharc} code.\cite{Mai2014SHARC}
\textsc{Sharc} is a generalization of the trajectory surface hopping method\cite{Tully1990JCP} that allows the simultaneous description of singlet and triplet states, coupled by nonadiabatic and spin-orbit couplings.\cite{Richter2011JCTC,Mai2015IJQC}

The initial conditions were sampled from a Wigner-Ville distribution\cite{Dahl1988JCP} of the ground state oscillator, based on MP2/ANO-RCC-VDZP harmonic frequencies.
For a set of 1000 initial geometries, vertical excitation calculations provided the initial electronic state distribution.\cite{Barbatti2007JPPA}
This procedure yielded 93~admissible initial conditions in the $5.90 \pm 0.15$~eV window around the absorption maximum of 5BU (at our level of theory).
Out of these, 82 trajectories were started in the $S_2$ and 11 in the $S_3$ state.

The trajectories were propagated for 400~fs, using a time step of 0.5~fs for the nuclear motion and 0.02~fs for the electronic wave function propagation.
The local diabatization procedure\cite{Granucci2001JCP} with wave function overlaps\cite{Plasser2016JCTC} was employed.
Decoherence was treated with an energy gap-based scheme.\cite{Granucci2007JCP}
From the total 93 trajectories, 15~trajectories were discarded due to convergence problems.

For each nuclear time step, the involved energies, gradients, wave function overlaps,\cite{Plasser2016JCTC} and spin-orbit couplings\cite{Mai2014JCP_reindex} were calculated at the multi-reference configuration interaction including single excitations (MRCIS) level of theory with the cc-pVDZ-DK basis set.\cite{Dunning1989JCP,Wilson1999JCP}
The COLUMBUS 7.0\cite{Lischka2012a,Lischka2011WCMS} program suite, employing integrals from MOLCAS,\cite{Aquilante2015JCC,Lischka2011WCMS} was used.
The orbitals were optimized using SA-CASSCF(10,8) (orbitals in Figure~\ref{fig:orbs}b), averaging simultaneously over 4~singlets and 3~triplet states.
In order to keep the computations feasible, the MRCIS reference space employed a restricted active space scheme, where in the reference configuration state functions only up to two electrons were allowed in the three antibonding orbitals.
This approach allowed retaining as many orbitals in the reference space as possible, to allow maximum flexibility in order to describe the three main deactivation pathways of this molecule.
This is particularly important here, since valence excitations and excitations responsible for the dissociation of the C-Br bond need to be considered on the same footing.


\section{Results and Discussion}

\subsection{Vertical Excitation Energies}

Table~\ref{tab:vertexc} compares the available experimental excitation energies of 5BU in gas phase\cite{Abouaf2003CPL} with the results of the vertical excitation calculations conducted here.

\begin{table}[ht]
  \caption{
  Comparison of experimental excitation energies (in eV) with vertical excitation calculations at the MS-CASPT2, CASSCF, MRCIS, ADC(2) and TD-DFT levels of theory.
  For all calculations, the CASSCF geometry from Ref.~\onlinecite{Kobylecka2009JPCA} was used.
  }
  \label{tab:vertexc}
  \centering
  \setlength{\tabcolsep}{1.3pt}
  \scriptsize
  \begin{tabular}{l l lll lll lll lll lll lll}
    \hline
            &Exp.\cite{Abouaf2003CPL}&\multicolumn{3}{l}{MS-CASPT2$^a$}&\multicolumn{3}{l}{CASSCF$^b$}  &\multicolumn{3}{l}{MRCIS$^c$}      &\multicolumn{3}{l}{ADC(2)$^d$}&\multicolumn{3}{l}{TD-BP86$^e$}&\multicolumn{3}{l}{TD-B3LYP$^f$} \\
    State   &&$E$    &$f_\text{osc}$ &Char.    &$E$    &$f_\text{osc}$ &Char.    &$E$    &$f_\text{osc}$ &Char.    &$E$    &$f_\text{osc}$ &Char.    &$E$    &$f_\text{osc}$ &Char.    &$E$    &$f_\text{osc}$ &Char.\\
    \hline
    $S_0$   &      &0.00   &---   &cs                                &0.00   &---   &cs                                &0.00   &---   &cs                                &0.00   &---   &cs                                &0.00   &---   &cs                                &0.00   &---   &cs                                \\
    $S_1$   &4.66  &4.60   &0.30  &$\pi_4\pi_1^*$                    &5.02   &0.00  &$n_\text{O1}\pi_1^*$              &5.09   &0.00  &$n_\text{O}\pi_1^*$               &4.75   &0.00  &$n_\text{O1}\pi_1^*$              &4.10   &0.00  &$n_\text{O1}\pi_1^*$              &4.66   &0.12  &$\pi_4\pi_1^*$                    \\
    $S_2$   &      &4.92   &0.00  &$n_\text{O1}\pi_1^*$              &6.17   &0.00  &$\pi_4\sigma_\text{CBr}^*$        &6.08   &0.29  &$\pi_4\pi_1^*$                    &4.95   &0.21  &$\pi_4\pi_1^*$                    &4.23   &0.08  &$\pi_4\pi_1^*$                    &4.78   &0.00  &$n_\text{O1}\pi_1^*$              \\
    $S_3$   &      &5.47   &0.00  &$\pi_4\sigma_\text{CBr}^*$        &6.76   &0.59  &$\pi_4\pi_1^*$                    &6.21   &0.00  &$\pi_4\sigma_\text{CBr}^*$        &5.64   &0.00  &$\pi_4\sigma_\text{CBr}^*$        &4.69   &0.00  &$n_\text{O2}\pi_2^*$              &4.98   &0.00  &$\pi_4\sigma_\text{CBr}^*$        \\
    $S_4$   &6.12  &6.01   &0.03  &$\pi_3\pi_1^*$                    &7.61   &0.01  &$\pi_3\pi_1^*$                    &7.74   &0.11  &$\pi_3\pi_1^*$                    &6.04   &0.00  &$n_\text{O2}\pi_2^*$              &4.73   &0.00  &$\pi_4\sigma_\text{CBr}^*$        &5.43   &0.00  &$n_\text{Br}\pi_1^*$              \\
    $S_5$   &      &6.23   &0.00  &$n_\text{O2}\pi_2^*$              &7.99   &0.00  &$n_\text{O2}\pi_2^*$              &8.13   &0.01  &$n_\text{O}\sigma_\text{CBr}^*$   &6.20   &0.04  &$\pi_3\pi_1^*$                    &4.87   &0.00  &$n_\text{Br}\pi_1^*$              &5.79   &0.00  &$n_\text{O2}\pi_2^*$              \\
    $S_6$   &      &6.55   &0.00  &$n_\text{Br}\sigma_\text{CBr}^*$  &8.52   &0.00  &$\pi_\text{Br}\sigma_\text{CBr}^*$&8.36   &0.00  &$n_\text{O}\pi_2^*$               &6.42   &0.00  &$n_\text{Br}\pi_1^*$              &5.01   &0.00  &$n_\text{O1}\sigma_\text{CBr}^*$  &5.88   &0.06  &$\pi_3\pi_1^*$                    \\
    $S_7$   &      &6.61   &0.00  &$n_\text{Br}\pi_1^*$              &8.66   &0.03  &$\pi_\text{Br}\pi_1^*$            &8.59   &0.14  &$\pi_4\pi_2^*$                    &6.64   &0.26  &$\pi_4\pi_2^*$                    &5.20   &0.05  &$\pi_3\pi_1^*$                    &5.90   &0.00  &$\pi_4\text{Ryd}$                 \\
    $S_8$   &7.20  &7.08   &0.00  &$\pi_\text{Br}\pi_1^*$            &8.83   &0.00  &$n_\text{O1}\sigma_\text{CBr}^*$  &8.64   &0.00  &$\pi_\text{Br}\sigma_\text{CBr}^*$&6.74   &0.00  &$n_\text{Br}\sigma_\text{CBr}^*$  &5.45   &0.09  &$\pi_4\pi_2^*$                    &6.08   &0.12  &$\pi_4\pi_2^*$                    \\
    \hline
    $T_1$   &3.35  &3.51   &---   &$\pi_4\pi_1^*$                    &3.81   &---   &$\pi_4\pi_1^*$                    &3.92   &---   &$\pi_4\pi_1^*$                    &3.63   &---   &$\pi_4\pi_1^*$                    &3.23   &---   &$\pi_4\pi_1^*$                    &3.20   &---   &$\pi_4\pi_1^*$                    \\
    $T_2$   &      &4.78   &---   &$n_\text{O1}\pi_1^*$              &4.88   &---   &$n_\text{O1}\pi_1^*$              &4.82   &---   &$n_\text{O}\pi_1^*$               &4.55   &---   &$n_\text{O1}\pi_1^*$              &3.89   &---   &$n_\text{O1}\pi_1^*$              &4.43   &---   &$n_\text{O1}\pi_1^*$              \\
    $T_3$   &      &5.01   &---   &$\pi_4\sigma_\text{CBr}^*$        &5.69   &---   &$\pi_3\pi_1^*$                    &5.68   &---   &$\pi_4\sigma_\text{CBr}^*$        &5.16   &---   &$\pi_4\sigma_\text{CBr}^*$        &4.35   &---   &$\pi_4\sigma_\text{CBr}^*$        &4.53   &---   &$\pi_4\sigma_\text{CBr}^*$        \\
    $T_4$   &      &5.16   &---   &$\pi_3\pi_1^*$                    &5.93   &---   &$\pi_4\sigma_\text{CBr}^*$        &6.31   &---   &$\pi_3\pi_1^*$                    &5.36   &---   &$\pi_3\pi_1^*$                    &4.51   &---   &$\pi_3\pi_1^*$                    &4.75   &---   &$\pi_3\pi_1^*$                    \\
    $T_5$   &      &5.91   &---   &$n_\text{Br}\sigma_\text{CBr}^*$  &6.83   &---   &$\pi_3\pi_2^*$                    &7.87   &---   &$\pi_\text{Br}\pi_1^*$            &5.90   &---   &$n_\text{O2}\pi_2^*$              &4.64   &---   &$n_\text{O2}\pi_2^*$              &5.20   &---   &$\pi_4\pi_2^*$                    \\
    $T_6$   &      &6.05   &---   &$\pi_4\pi_2^*$                    &7.80   &---   &$n_\text{O2}\pi_2^*$              &8.00   &---   &$n_\text{O}\sigma_\text{CBr}^*$   &6.00   &---   &$n_\text{Br}\sigma_\text{CBr}^*$  &4.73   &---   &$\pi_4\pi_2^*$                    &5.37   &---   &$n_\text{Br}\pi_1^*$              \\
    $T_7$   &      &6.24   &---   &$n_\text{O2}\pi_2^*$              &8.02   &---   &$\pi_3\sigma_\text{CBr}^*$        &8.05   &---   &$\pi_\text{Br}\sigma_\text{CBr}^*$&6.03   &---   &$\pi_4\pi_2^*$                    &4.76   &---   &$n_\text{Br}\pi_1^*$              &5.56   &---   &$n_\text{Br}\sigma_\text{CBr}^*$  \\
    $T_8$   &      &6.99   &---   &$n_\text{Br}\pi_1^*$              &8.34   &---   &$\pi_4\pi_2^*$                    &8.24   &---   &$n_\text{O}\pi_2^*$               &6.38   &---   &$n_\text{Br}\pi_1^*$              &4.94   &---   &$n_\text{Br}\sigma_\text{CBr}^*$  &5.58   &---   &$n_\text{O2}\pi_2^*$              \\
    \hline
    \multicolumn{2}{l}{RMSD}    &0.00&&&1.01&&&0.75&&&0.21&&&0.67&&&0.34\\
    \hline
  \end{tabular}

  \raggedright

  Orbital labels refer (approximately) to Figure~\ref{fig:orbs}a, except for MRCIS (those refer to Figure~\ref{fig:orbs}b).
  $^a$ MS-CASPT2(20,14)/ANO-RCC-VDZP.  $S_0$ energy: --3017.895007~a.u.
  $^b$ CASSCF(14,11)/cc-pVDZ-DK.  $S_0$ energy: --3016.429666~a.u.
  $^c$ MRCIS(10,8)/cc-pVDZ-DK.  $S_0$ energy: --3016.550517~a.u.
  $^d$ ADC(2)/def2-TZVP.  $S_0$ energy: --2986.073460~a.u.
  $^e$ TD-BP86/TZP.  $S_0$ binding energy: --2.845430~a.u.
  $^f$ TD-B3LYP/TZP.  $S_0$ binding energy: --3.361766~a.u.
\end{table}

The singlet and triplet $\pi\pi^*$ excitations were estimated experimentally to be 4.66 eV\cite{Abouaf2003CPL} and 3.35~eV,\cite{Abouaf2003CPL,Rothman1967PP} respectively.
These excitations are accurately described by the $S_1$ and $T_1$ states of our MS-CASPT2 computations, with energies of 4.60  and 3.51~eV, respectively.
The presence of the dark singlet $n_\text{O1}\pi^*$ ($S_2$) state (in the following, if not stated otherwise ``$n_\text{O}\pi^*$'' refers to states involving the $n_\text{O1}$ orbital) energetically close to the spectroscopically active $^1\pi\pi^*$ state is reminiscent of the situation encountered in T and U.\cite{Zechmann2008JPCA,Asturiol2010PCCP,Yamazaki2012JPCA}
The third excited state, $S_3$, is an excitation into the antibonding orbital of the C-Br bond ($\pi_4\sigma_\text{CBr}^*$) and thus a state related to C-Br dissociation.
The $S_4$ is another $\pi\pi^*$ state, located at 6.01~eV.
The order of the singlet states agrees with that obtained with previous calculations on 5BU.\cite{Kobylecka2009JPCA, Storoniak2011JPCB}
The character of the triplet states mirror the singlet counterparts up to $T_4$, with the lowest triplet state being the $^3\pi\pi^*$ state, followed by the $^3n_\text{O}\pi^*$ state ($T_2$).

Due to their good agreement with the experimental values, the MS-CASPT2 values serve as reference data for the other levels of theory.
As an estimator for the accuracy of the excited-state calculations, we computed root mean square deviations (RMSD, bottom of Table~\ref{tab:vertexc}), comparing states of matching characters against the MS-CASPT2 results.
For the RMSD calculation we only included the relevant, low-lying states with $\pi_4\pi^*_1$, $n_\text{O1}\pi^*_1$, or $\pi_4\sigma^*_\text{CBr}$ characters---which are the $S_{1-3}$ or $T_{1-3}$ for MS-CASPT2 but might be different adiabatic states for the other methods.
Among the five cheaper levels of theory, the worst result is obtained with CASSCF (1.01~eV RMSD) and the best result with ADC(2) (0.21~eV).
The CASSCF and MRCIS methods both significantly overestimate the energies of the $^1\pi\pi^*$ states, changing the order of singlet excited states; this observation is due to an imbalanced description of correlation energy in these methods.
TD-BP86 systematically underestimates all excitation energies by about 1~eV on average, and alters the order of the states as well.
On the other hand, TD-B3LYP describes the excitation energies fairly well, in particular the bright $^1\pi\pi^*$ state, and gives the correct order for the three lowest singlet states.
In general, only the few lowest excited states in each multiplicity are consistently obtained with all methods, although the ordering can vary; for the higher states, different characters were obtained, and the energy deviations are larger than for the lower states.


\subsection{Potential Energy Surfaces}

The calculated vertical excitations are only the first step in disentangling the actual photophysical relaxation pathways of 5BU.
Hence, as a second step we investigated the most important pathways---ground state relaxation, ISC, and photodissociation---in more detail.
Because of the high cost of MS-CASPT2(20,14), the necessary optimizations were carried out with MS-CASPT2(12,9), which excludes the $n_\text{Br}$, $\pi_\text{Br}$, $\sigma_\text{CBr}$, $\sigma_\text{CBr}$, and $n_\text{O2}$ orbitals.
This reduced active space is only suitable to describe the lowest excitations (the $\pi\pi^*$ and $n_\text{O1}\pi^*$ states), but allows optimizing the critical points for these states.
As ADC(2) gave very promising results in the Franck-Condon region, we also performed these computations with ADC(2), to see whether its accuracy extends to the rest of the PESs.

\begin{figure}
  \centering
  \includegraphics[scale=1]{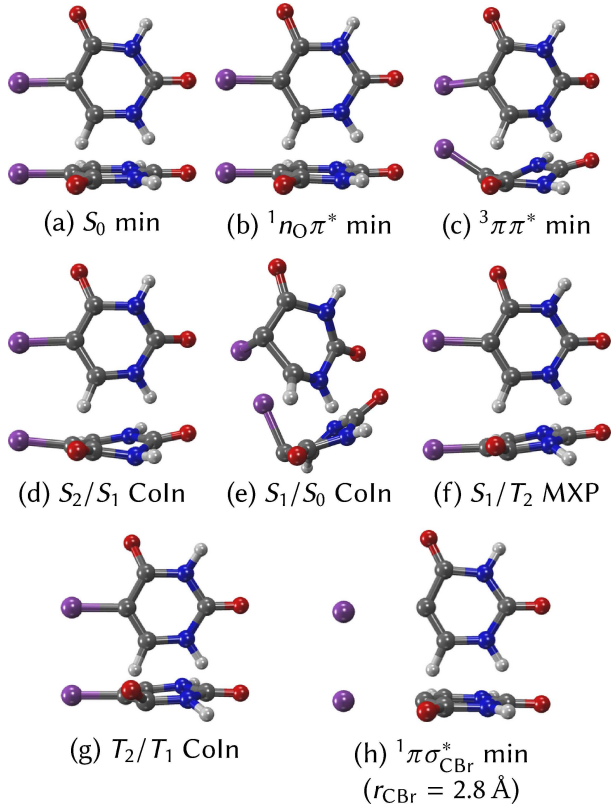}
  \caption{
  Geometries of the critical points optimized at the MS-CASPT2(12,9) level of theory.
  }
  \label{fig:geoms}
\end{figure}

The following minima were obtained at both levels of theory: the $S_0$ minimum, a minimum of the $^1n_\text{O}\pi^*$ state (located on the $S_1$ adiabatic surface), and a $^3\pi\pi^*$ minimum ($T_1$), shown in Figure~\ref{fig:geoms} a-c.
Additionally, crossing points between several electronic states (adiabatic surfaces in parenthesis) were found: a $^1\pi\pi^*/{}^1n_\text{O}\pi^*$ ($S_2/S_1$) conical intersection (CoIn), a $^1\pi\pi^*/S_0$ ($S_1/S_0$) CoIn, a $^1n_\text{O}\pi^*/{}^3\pi\pi^*$ ($S_1/T_2$) minimum-energy crossing (MXP), and a $^3\pi\pi^*/{}^3n_\text{O}\pi^*$ ($T_2/T_1$) CoIn, which are presented in Figure~\ref{fig:geoms} d-g.
The attempted optimization of a $^1\pi\pi^*$ minimum ended in the $^1\pi\pi^*/S_0$ CoIn.
We also optimized the geometry of the uracilyl radical in the $S_1$ ($^1\pi\sigma^*$) state Figure~\ref{fig:geoms} h; this optimization was constrained to $r_\text{CBr}=2.8\AA$ and $\alpha_\mathrm{C_4C_5Br}=119^\circ$, in order to allow for sensible results with ADC(2).
All optimized geometries can be found in the Supporting Information.
Based on the geometries, the following three chains of LIIC scans were generated:
(i) $S_0$ min -- $S_2/S_1$ CoIn (only ADC(2)) -- $S_1/S_0$ CoIn;
(ii) $S_0$ min -- $S_2/S_1$ CoIn -- $S_1$ min -- $S_1/T_2$ MXP -- $T_2/T_1$ CoIn -- $T_1$ min;
(iii) $S_0$ min -- constrained $S_1$ min.

\begin{figure*}
  \centering
  \includegraphics[scale=1]{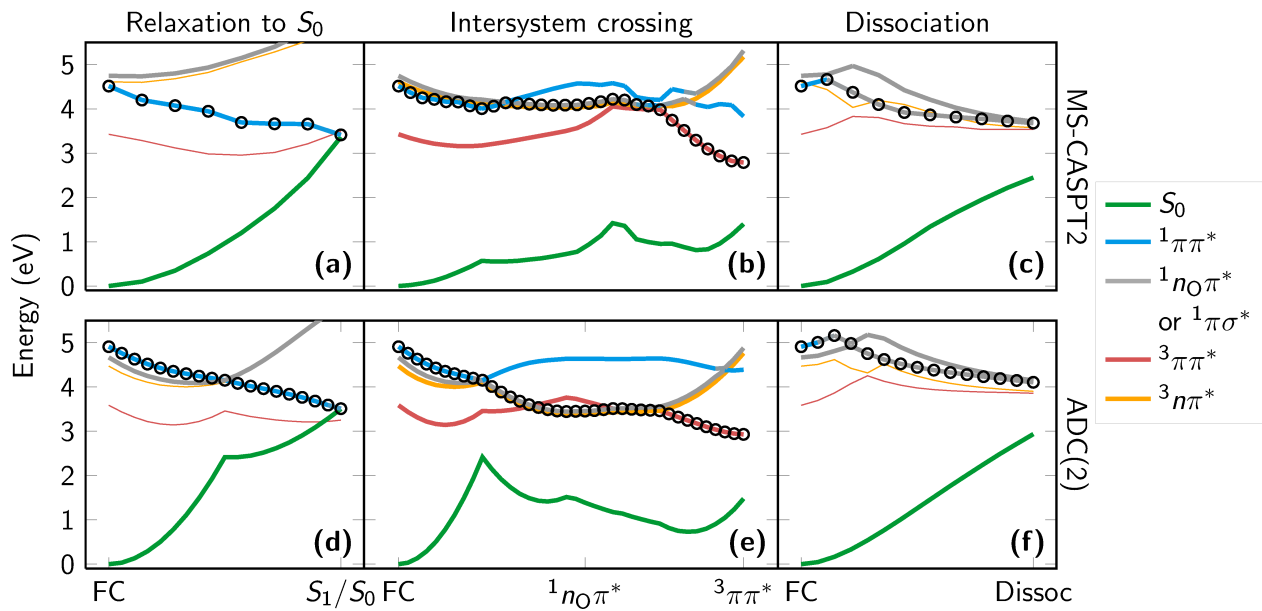}
  \caption{
  Linear interpolation in internal coordinates (LIIC) scans for the most important photophysical paths in 5BU, using the MS-CASPT2(20,14)//MS-CASPT2(12,9) level of theory (a-c), and ADC(2) (d-f).
  }
  \label{fig:liic}
\end{figure*}

The energies along these LIIC paths were recomputed with MS-CASPT2(20,14) and ADC(2), yielding the curves presented in Figure~\ref{fig:liic}.
At MS-CASPT2 level, ground state relaxation (Figure~\ref{fig:liic}~a) can be achieved by a barrierless, descending path from the Franck-Condon region to the $S_1/S_0$ CoIn.
This CoIn is related to a molecular geometry with strong ring puckering (depicted in Figure~\ref{fig:geoms}~e), as already reported for other pyrimidine nucleobases.\cite{Merchan2013TCC,Yamazaki2012JPCA}
As the bright state is the $S_1$ at this level, no other state crossing is involved, which should make ground state relaxation an important and efficient relaxation route.
Noteworthy, our calculations do not show the barrier reported by Koby\l{}ecka et al.,\cite{Kobylecka2009JPCA} probably because they employed CASSCF-based minimum energy paths, whereas we performed optimizations at the higher MS-CASPT2 level of theory.

In Figure~\ref{fig:liic}~b, it can be see that ISC in 5BU involves first switching from the bright $^1\pi\pi^*$ state to the $^1n_\text{O}\pi^*$ state.
Subsequently, close to the nearly planar $^1n_\text{O}\pi^*$ minimum (Figure~\ref{fig:geoms}~b) a near-degeneracy region with the $^3\pi\pi^*$ state can be reached through a slight pyramidalization at carbon atom C$_4$ (Figure~\ref{fig:geoms} f and g).
Note that in this region, the lowest triplet states are exchanged, with $^3\pi\pi^*$ being $T_2$, unlike at the Franck-Condon geometry.
The spin-orbit coupling between the $^1n_\text{O}\pi^*$ and $^3\pi\pi^*$ states amounts to approximately 60~cm$^{-1}$ at this point, allowing for ISC.
Decay to the lowest triplet minimum ($^3\pi\pi^*$), which shows a boat-like ring conformation (Figure~\ref{fig:geoms}~c) like in uracil and thymine,\cite{Serrano-Perez2007JPCB} is possible through the nearby $T_2/T_1$ CoIn.
The energy of the $^3\pi\pi^*$, 2.9~eV, fits very well with the 3.0~eV onset of the phosphorescence spectrum of 5BU.\cite{Rothman1967PP}
The ISC route in 5BU is reminiscent of the one found in U\cite{Richter2014PCCP} and T,\cite{Mai2016CP} where also the $^1n_\text{O}\pi^*$ minimum is the gateway for population of the triplet manifold, due to the presence of the nearby $S_1/T_2$ MXP and $T_2/T_1$ CoIn (compare Figures~\ref{fig:geoms}~b, f, and g).
Also of interest is the observation that from the $^1n_\text{O}\pi^*$ minimum the system could easily cross back to the $^1\pi\pi^*$ state and subsequently deactivate to the ground state, as was already proposed in the literature for U and T.\cite{Mercier2008JPCB,Yamazaki2012JPCA}

We note that the spin-orbit couplings in 5BU are not much stronger than in U or T, although one might have expected this from the heavy-atom effect of the bromine atom.
The relatively weak spin-orbit couplings can be explained by the fact that neither the $^1n_\text{O}\pi^*$ nor the $^3\pi\pi^*$ excitations involve significantly the bromine orbitals.
Due to the properties of the spin-orbit Hamiltonian, very large spin-orbit couplings could only be expected if the two involved states differ in one orbital localized at the heavy bromine atom.
For example, the coupling between the $^1\pi_4\sigma_\text{CBr}^*$ and $^3n_\text{Br}\sigma_\text{CBr}^*$ states amounts to 860~cm$^{-1}$ at the Franck-Condon geometry.
However, these strong interactions are only relevant at higher energies, but do not play a role after excitation to the $S_1$.
Hence, in the low-energy range, ISC in 5BU should not be significantly different than ISC in U.

In Figure~\ref{fig:liic}~c, the potential energy curves suggest that in order to initiate photodissociation, the system would need to switch from the bright $^1\pi\pi^*$ state to the dissociative $^1\pi\sigma^*$ state.
At least in the shown scan, this involves a barrier of 0.2~eV.
Moreover, the barrier might become larger after the molecule relaxes in the $^1\pi\pi^*$ state.
In this regard, a transition state search based on MS-CASPT2 potentials would be helpful.
Unfortunately, the minimum energy path presented in Ref.~\onlinecite{Kobylecka2009JPCA} is not, since this path was optimized using CASSCF(12,10), with an active space that excludes the $n_\text{O}$ orbitals.
This makes the dissociative $\pi\sigma^*$ state the lowest state in their optimization (according to our reproduction of their calculation), explaining why their minimum-energy path was dissociative.
Instead, we expect that a minimum-energy path using the $^1\pi\pi^*$ gradients (e.g., at the MS-CASPT2 level) would not show dissociation, but rather approaches the $S_1/S_0$ CoIn.

Dissociation could in principle also happen from the $^1n_\text{O}\pi^*$ minimum or via the triplet states, as experimentally observed in solution by Swanson et al.\cite{Swanson1981JACS}
The $^1n_\text{O}\pi^*$ minimum has a higher energy than the dissociation limit (according to Figure~\ref{fig:liic}~c), but we expect that there exists some barrier due to the avoided crossing between $^1n_\text{O}\pi^*$ and the dissociative states.
Dissociation in the triplet state should be hampered by the low energy of the $T_1$ minimum (0.7~eV below the dissociation limit), and hence would need to occur from an unrelaxed triplet.

The bottom row of Figure~\ref{fig:liic} (panels d, e, and f) presents the equivalent paths obtained from the ADC(2)/def2-TZVP calculations.
For ground state relaxation (Figure~\ref{fig:liic}~d), the path involves the $S_2/S_1$ CoIn, because the bright state at ADC(2) level is wrongly the $S_2$.
This state reordering could increase the probability to switch to the $^1n_\text{O}\pi^*$ state during decay, but otherwise, ADC(2) agrees with MS-CASPT2 in that the ground state relaxation path is barrierless.

Regarding the ISC path (Figure~\ref{fig:liic}~e), the most important feature of the ADC(2) curves is the low energy of the $^1n_\text{O}\pi^*$ minimum relative to the $S_2/S_1$ crossing (in contrast to the MS-CASPT2 result, where the minimum is only 0.1~eV below the crossing).
At the ADC(2) level, this would probably lead to population trapping in the $^1n_\text{O}\pi^*$ minimum with subsequent ISC, whereas recrossing to the $^1\pi\pi^*$ state would be strongly suppressed.
However, ISC itself seems to be well described at the ADC(2) level, with the $^1n_\text{O}\pi^*/{}^3\pi\pi^*$ crossing energetically close to the $^1n_\text{O}\pi^*$ minimum.

Finally, Figure~\ref{fig:liic}~f shows the potential curves for photodissociation at ADC(2) level of theory.
Because ADC(2) is a single-reference method and cannot properly describe the full dissociation, we only present the energy scan for the C-Br bond length from 1.9~\AA\ up to 2.8~\AA\ (see Figure~\ref{fig:geoms}~h).
Despite the restricted scan, one can still study the barrier between the $^1\pi\pi^*$ and $^1\pi\sigma^*$ states, and obtain an estimate of the dissociation energy.
From the scan in Figure~\ref{fig:liic}~f, it can be seen that ADC(2) qualitatively agrees with MS-CASPT2, giving a barrier of 0.2~eV, which is due to the avoided crossing between $^1\pi\pi^*$ and $^1\pi\sigma^*$.
Because at the ADC(2) level the $^1n_\text{O}\pi^*$ and triplet minima are quite low in energy, dissociation would be even less likely than at MS-CASPT2 level.

Based on the MS-CASPT2 results, we can therefore propose that ground state relaxation should be the predominant relaxation route of 5BU in gas phase.
However, the close proximity of the bright $^1\pi\pi^*$ state to the $^1n_\text{O}\pi^*$ state could lead to some population transfer into the latter state.
Due to the favorable location of the $^1n_\text{O}\pi^*/{}^3\pi\pi^*$ crossing, it is then conceivable that some ISC occurs.
Dissociation is disfavored due to the barrier needed to overcome before populating the dissociative $^1\pi\sigma^*$ or $n_\text{O}\sigma^*$ states.
These findings do not fully agree with the conclusions drawn by Blancafort and coworkers,\cite{Kobylecka2009JPCA} who proposed that in 5BU homolytic C-Br cleavage and relaxation to the ground state are in close competition, based on SA-CASSCF calculations.

\subsection{Choice of Electronic Structure Method for Dynamics}

While the presented static quantum chemistry calculations reveal some new details on the excited-state behavior of 5BU, it would be also advantageous to perform excited-state dynamics simulations, to obtain excited-state lifetimes or branching ratios between the relaxation channels.

Ideally, dynamics simulations in this case should be carried out with our reference method MS-CASPT2(20,14); however, this method is clearly too expensive for dynamics.
Even with the smaller active space CAS(12,9) employed for the optimizations, computing nuclear gradients for MS-CASPT2 remains very costly.
Moreover, TD-DFT has been known to severely overestimate the energies of non-planar geometries in nucleobases.\cite{Barbatti2012JCP}
For 5BU, we attempted to optimize the $S_1/S_0$ CoIn using the two TD-DFT methods from Table~\ref{tab:vertexc}, and found energy barriers of at least 2~eV from the Franck-Condon region to the CoIn, showing the inadequacy of these methods.

Currently, the only practical choices to run \textsc{Sharc} simulations are CASSCF and MRCIS.
The shared weakness of both methods is their significant overestimation of the $^1\pi\pi^*$ energies, which in this particular case can prevent ground state relaxation and enhance dissociation.
From Table~\ref{tab:vertexc} it is clear that this overestimation is even more severe in CASSCF than in MRCIS.
Whereas at CASSCF level, the bright $^1\pi\pi^*$ state is 0.6~eV above the dissociative $^1\pi\sigma^*$ state, MRCIS predicts $^1\pi\pi^*$ to be 0.1~eV below $^1\pi\sigma^*$.
Furthermore, as reviewed in Ref.~\onlinecite{Mai2014TCC}, for U and T CASSCF predicts significant trapping in the $^1\pi\pi^*$ state, which is also not consistent with MS-CASPT2 calculations.\cite{Yamazaki2012JPCA}
Hence, while both MRCIS and CASSCF do not fully agree with the MS-CASPT2 results, MRCIS is still clearly the preferable choice among those two options.

\begin{figure}
  \centering
  \includegraphics[scale=1]{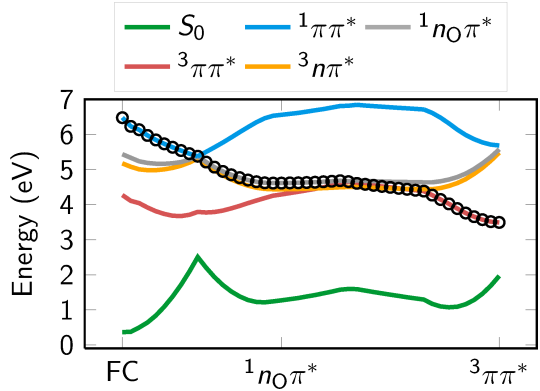}
  \caption{
  Linear interpolation in internal coordinates (LIIC) scans as in Figure~\ref{fig:liic}b and e, but for the MRCIS(10,8) level of theory.
  }
  \label{fig:liic_mrcis}
\end{figure}

In order to investigate which parts of the PESs of 5BU can be described reasonably at the MRCIS level of theory, we also performed the LIIC calculations with this method.
As expected, dissociation is not properly described (not shown), with a nearly barrierless path from the Franck-Condon region to the dissociation limit.
Furthermore, due to the high energy of the $^1\pi\pi^*$ state, the CoIn between $S_0$ and $^1\pi\pi^*$ is not as accessible as it should be, and relaxation to the ground state is hampered.
The ISC pathway appears to be qualitatively correct.
As can be seen in Figure~\ref{fig:liic_mrcis}, the ISC pathway at MRCIS level is very similar to the one at ADC(2) level.
In particular, after excitation to the bright $^1\pi\pi^*$ ($S_2$) state, the system first needs to cross from $^1\pi\pi^*$ to $^1n_\text{O}\pi^*$ at the corresponding $S_2/S_1$ CoIn and then relax to the $^1n_\text{O}\pi^*$ minimum.
There, a barrier of 0.1~eV needs to be surmounted to reach a crossing with the $^3\pi\pi^*$ (which is the adiabatic $T_2$ state at this point) with SOCs of about 55~cm$^{-1}$, allowing ISC to take place.
Once transferred to the $T_2$ state, the system can relax through the $T_1/T_2$ CoIn to the $^3\pi\pi^*$ minimum, located on the adiabatic $T_1$.
Hence, we conclude that trajectories at the MRCIS level of theory will be able to properly sample the ISC pathway of 5BU, although unfortunately they will only provide qualitative aspects of the deactivation mechanism, because the ground state relaxation and dissociation pathways are under- and over-represented, respectively.

\subsection{Absorption Spectrum}

\begin{figure}
  \centering
  \includegraphics[scale=1]{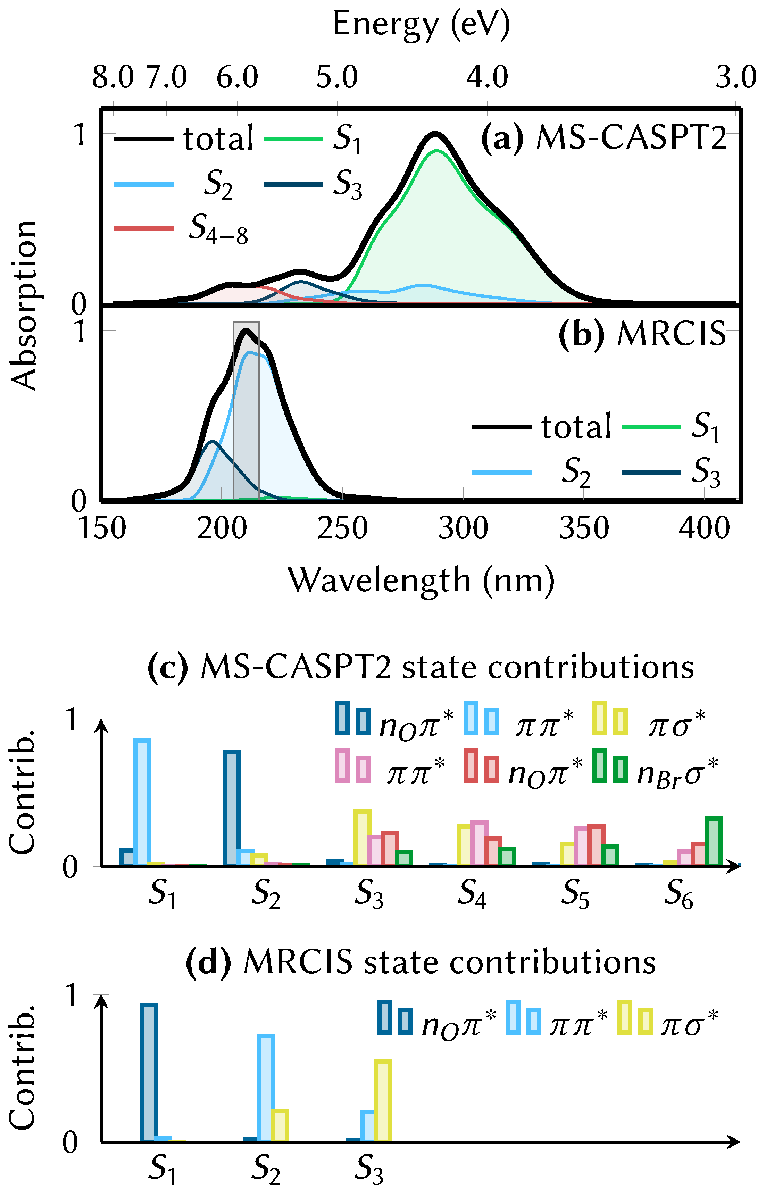}
  \caption{
  Calculated absorption spectrum of the \emph{keto} tautomer of 5BU (a) at the MS-CASPT2(20,14)/ANO-RCC-VDZP and (b)  MRCIS(10,8)/cc-pVDZ-DK levels of theory.
  The solid black lines show the total absorption spectrum; color lines indicate the contributions of the different singlet states, as indicated.
  Grey box in (b) denotes the excitation window employed in the dynamics simulations.
  In (c) and (d), the contributions of the state characters to the absorption bands in (a) and (b) are given.
  }
  \label{fig:spec}
\end{figure}

In order to prepare the initial conditions for the excited-state dynamics, we first simulated the absorption spectrum of 5BU, based on a distribution of geometries representing the ground state vibrational state.
In Figure~\ref{fig:spec}a,b we show the simulated spectrum, as obtained with MS-CASPT2(20,14) and MRCIS, respectively.
As expected from the single point calculation at the equilibrium geometry, the MS-CASPT2 spectrum (depicted in Figure~\ref{fig:spec}a) shows an absorption band centered at approximately 4.3~eV, due to the bright $\pi\pi^*$ ($S_1$) state.
The character of the involved states, shown in Figures \ref{fig:spec}c, was analyzed by calculating wave function overlaps to the states at the equilibrium geometry; these overlaps allow to compare states for hundreds of calculations without individually inspecting the orbital and expansion coefficients.\cite{Plasser2016JCP}
This analysis showed that for the large majority of sampled geometries (Figure~\ref{fig:spec}c), $S_1$ has $\pi\pi^*$ character and $S_2$ has $n_\text{O}\pi^*$ character, whereas for the higher states the character varies with geometry, e.g., the $\pi\sigma^*$ state which is distributed between the $S_2$-$S_5$.
Comparing the spectrum to experimental spectra, we find that the MS-CASPT2 calculations qualitatively reproduce the first absorption band of 5BU, but the sampling has shifted the peak to slightly lower energies than those predicted experimentally.\cite{Abouaf2003CPL,Dietz1987JACS,Lu2004JPCB}

The MRCIS spectrum, shown in Figure~\ref{fig:spec}b, presents a peak centered at 5.9~eV, arising from the spectroscopically active $\pi\pi^*$ state, which is either $S_2$ or $S_3$ depending on the initial geometry.
This peak is shifted by 1.6~eV with respect to the MS-CASPT2 spectrum, as expected from the vertical excitation calculations (see Table~\ref{tab:vertexc}).
At almost all geometries, $S_1$ is the dark $^1n_\text{O}\pi^*$ state (Figure~\ref{fig:spec}d), explaining why $S_1$ does not contribute to the spectrum.
Instead, the absorption band is dominated by the bright $^1\pi\pi^*$ state, which is distributed between the $S_2$ and $S_3$ states. 
Contributions to the $\pi\sigma^*$ dissociative state can also be found in $S_2$ and $S_3$.

As with the vertical excitation energies, the comparison of the MRCIS spectrum with experimental data\cite{Abouaf2003CPL,Dietz1987JACS,Lu2004JPCB} suffers from the major drawback that the energy of the $\pi\pi^*$ transition is blue-shifted at the MRCIS(10,8)/cc-pVDZ-DK level of theory.
However, we note that for the majority of initial conditions, the bright $^1\pi\pi^*$ state is below the dissociative $^1\pi\sigma^*$ state.
This should limit the amount of artificial dissociation so that still a number of trajectories can sample the other relaxation pathways, in particular ISC.


\subsection{Evolution of the Trajectories}

We shall start the discussion of the nonadiabatic dynamics data with the results obtained from an analysis of the kinetic constants.
As a general finding, about a third of the propagated trajectories (23 of 78) undergo C-Br bond dissociation within 400~fs. Clearly, this dissociation quantum yield is farfetched in comparison with the experimental few-percent yields,\cite{Campbell1974ZNB,Swanson1981JACS,Dietz1987JACS}
due to the level of theory employed in the dynamics.
From the highly accurate MS-CASPT2 results, no ultrafast dissociation should be expected in gas phase after $^1\pi\pi^*$ excitation.
Therefore, we do not further discuss those trajectories that dissociated; only the 55~non-dissociating trajectories are included in the following analysis.

\begin{figure}
  \centering
  \includegraphics[scale=1]{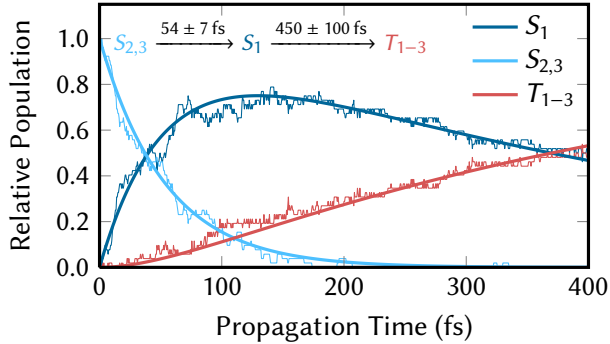}
  \caption{
  Temporal evolution of excited-state populations of the 55~non-dissociating trajectories, computed with \textsc{Sharc} and MRCIS(10,8)/cc-pVDZ-DK.
  }
  \label{fig:pop}
\end{figure}

Figure~\ref{fig:pop} presents the temporal evolution of the excited-state populations and the results of the kinetic analysis.
The populations were fitted to a sequential kinetic model with two parameters $\tau_1$ and $\tau_2$, which we identify as the internal conversion (IC) and ISC time constants, respectively.
Errors for the parameters were obtained with the bootstrapping method,\cite{Nangia2004JCP} in order to judge whether the number of trajectories is sufficient to describe these processes.

We obtain $\tau_1=54\pm 7$~fs, meaning that the $^1\pi\pi^*\rightarrow{} ^1n_\text{O}\pi^*$ population transfer is about 70\% after 100~fs (see the $S_1$ line in Figure~\ref{fig:pop}).
Note that this is substantially different to the results reported for uracil and thymine in previous SA-CASSCF-based dynamics simulations, where much less was transferred within 100~fs, because longer time constants were obtained for this pathway.\cite{Hudock2007JPCA,Szymczak2009JPCA,Nachtigallova2011JPCA,Fingerhut2013JPCL,Richter2014PCCP,Mai2016CP}
This difference is due to the fact that at SA-CASSCF level the $^1\pi\pi^*$ state is higher in energy than with MRCIS, and therefore population transfer from $^1\pi\pi^*$ to $^1n_\text{O}\pi^*$ is slower at SA-CASSCF level.

The second time constant we obtain is $\tau_2=450\pm 100$~fs.
That is, within 400~fs, 51\% of the non-dissociating trajectories have moved to the triplet states and 49\% are still in the singlet manifold.
The calculated ISC time scale matches well with the 0.4~ps time constant reported in Ref.~\onlinecite{Wang2006JCPa} for time-resolved pump-probe experiments (266~nm pump and 330~nm probe) on 5BU in water.
In the same experiment, they also observe a nanosecond component, which could be assigned to the population trapped in the triplet states.

Interestingly, the time constant of 450~fs obtained for ISC in 5BU is smaller than those previously reported for U and T,\cite{Richter2014PCCP,Mai2016CP} which are approximately 800~fs and 900~fs, respectively.
The factor of two in the ISC time constant is due to subtle differences in the excited-state PESs, due to the presence of the bromine atom, and ultimately because a different level of theory has been used in the previous simulations.\cite{Richter2014PCCP,Mai2016CP}
Otherwise, the general ISC mechanism in 5BU, T, and U is the same and also the spin-orbit couplings have approximately the same size.
However, while the ISC time constant appears to be reasonable, the overall triplet yield is clearly overestimated, because a large fraction of the trajectories should undergo relaxation to $S_0$ before ISC can occur.
In our simulations, this relaxation is blocked by the large barrier to recross back to the $^1\pi\pi^*$ and to reach the $^1\pi\pi^*/S_0$ CoIn.
Consequently, for the simulated time, the MRCIS-based dynamics does not result in trajectories undergoing relaxation to the ground state, as was suspected from the static calculations.

These results show that a relevant deactivation pathway of 5BU should be ultrafast ISC.
In particular, the results of the interpolation scans in Figure~\ref{fig:liic} and \ref{fig:liic_mrcis} are reminiscent to the situation reported for U and T,\cite{Matsika2004JPCA,Richter2014PCCP,Serrano-Perez2007JPCB} where also from the bright $^1\pi\pi^*$ state the system can deactivate to the dark $S_1$ ($n_\text{O}\pi^*$) state.
This state is the doorway for ISC into the $^3\pi\pi^*$ state, in accordance with the El-Sayed rules.\cite{El-Sayed1963JCP}
For 5BU, this ISC mechanism is in agreement with experimental data,\cite{Dietz1987JACS,Swanson1981JACS} which report ISC from an $^1n_\text{O}\pi^*$ to a $^3\pi\pi^*$ state.
The involved geometries differ from the ground state geometry by compressions and elongations along the C=O bonds and deformations of the ring.


\section{Conclusions}

We performed accurate vertical excitation calculations and potential energy surface explorations for 5-bromouracil (5BU), in order to investigate the possible excited-state dynamics of this molecule.
The vertical excitation calculations showed that MS-CASPT2 with an active space of 20~electrons in 14~orbitals gives excitation energies in excellent agreement with experiment.
In particular, the lowest excited states are the bright $^1\pi\pi^*$ state, a dark $^1n_\text{O}\pi^*$ state, and a dissociative $^1\pi\sigma^*$ state at slightly higher energies.
These three states are responsible for the three plausible photophysical reaction pathways of 5BU, according to MS-CASPT2.
A barrierless pathway on the $^1\pi\pi^*$ state leads directly to a conical intersection with the $S_0$, allowing for efficient relaxation to the ground state.
Alternatively, another conical intersection allows the interconversion of $^1\pi\pi^*$ into $^1n_\text{O}$, from where intersystem crossing to the $^3\pi\pi^*$ state can commence.
Third, after surmounting a barrier of at least 0.2~eV, the molecule could switch from $^1\pi\pi^*$ to $^1\pi\sigma_\text{CBr}^*$, which would lead to the homolytic dissociation of the C-Br bond.
Out of these three pathways, MS-CASPT2 identifies relaxation to $S_0$ as the most probable pathway in gas phase.

We also investigated the performance of different electronic structure methods in describing the deactivation of 5BU, including  ADC(2), TD-BP886, TD-B3LYP, SA-CASSCF, and MRCIS.
All of these methods represent significantly lower-cost alternatives to the expensive MS-CASPT2 computations performed.
From the excitation energies and potential energy surfaces, ADC(2) seems to give the best results in agreement with the MS-CASPT2 reference calculations.
The DFT-based methods fail especially in describing the ground state relaxation pathway.
SA-CASSCF and MRCIS, on the other hand, notably overestimate the energy of the $^1\pi\pi^*$ state, which makes dissociation more likely and ground state relaxation less likely.
For technical constraints, MRCIS was chosen to perform nonadiabatic dynamics simulations coupled to \textsc{Sharc}, a variant of the trajectory surface hopping method that can treat intersystem crossing.
We found that intersystem crossing is mediated by the $^1n_\text{O}\pi^*$ state and occurs with a time constant of $450\pm100$~fs, showing that this process is ultrafast in 5BU, as already discussed in recent literature on pyrimidine bases.\cite{Mai2014TCC,Richter2014PCCP,Mai2016CP,Etinski2009JPCA}
Unfortunately, the trajectories do not allow obtaining yields for the competitive photophysical pathways of 5BU, due to the shortcomings of MRCIS.

The present study clearly shows that the proper description of the excited-state dynamics of 5BU is very challenging, and requires a multiconfigurational electronic structure method with a large active space, in order to account for all different, competing relaxation pathways.
Moreover, previous experiments were conducted in condensed phase, and the applications of 5BU involve the photoactive molecule in a complex biological environment, and the inclusion of such environments would add another level of complexity to future computations.


\section*{Competing Interests}

The authors declare no competing interests.


\section*{Author Contributions}

F.P.\ performed the CASSCF and MRCIS vertical excitations calculations, the \textsc{Sharc} dynamics simulations and drafted the manuscript.
S.M.\ performed the rest of the quantum chemistry calculations, provided support for the dynamics simulations, performed the dynamics analysis and contribute to the draft.
L.G.\ conceived, supervised the study and contributed to the writing of the manuscript.
All authors discussed the results and approved the manuscript for publication.


\section*{Acknowledgements}

The authors like to thank Philipp Marquetand and Mariona Sodupe for fruitful discussions. 
The Cost Action ECostBio CM1305 is also acknowledged for stimulating discussions.
F.P.\ acknowledges the Universitat Aut\`onoma de Barcelona for a PIF scholarship that enabled her research stay at the University of Vienna.

\section*{Funding}

This work is funded by the Austrian Science Fund (FWF) project P25827, the University of Vienna and the Universitat Aut\'onoma de Barcelona. 
Part of the calculations have been performed at the Vienna Scientific Cluster (VSC).



\clearpage
\linespread{1.0}
\begin{center}
{
\Large Supporting Information for:\\Insights into the Deactivation of 5-Bromouracil after UV Excitation
}
{\small
\\
\vspace{0.3cm}
Francesca Peccati,$^a$ Sebastian Mai,$^b$ Leticia Gonz\'alez$^{b,*}$\\
\vspace{0.3cm}
$^{a}$ Departament de Qu\'{\i}mica, Universitat Aut\`onoma de Barcelona, 08193 Bellaterra, Spain.\\
$^{b}$ Institute of Theoretical Chemistry, Faculty of Chemistry, University of Vienna, Währinger Str. 17, 1090 Vienna, Austria.\\
\vspace{0.3cm}
$^*$ Corresponding author\\
\textit{Email address:} leticia.gonzalez@univie.ac.at (Leticia Gonz\'alez)\\
}

\end{center}



\setcounter{section}{0}
\setcounter{figure}{0}
\setcounter{table}{0}
\renewcommand{\thefigure}{S\arabic{figure}}
\renewcommand{\thetable}{S\arabic{table}}

\section*{MS-CASPT2 Geometries}

The following geometries of 5-bromouracil were optimized with MS-CASPT2(12,9)/ANO-RCC-VDZP.
For state-averaging and multi-state treatment, four singlets or three triplets were used.
The calculations were done with \textsc{Molcas} 8.0, using an IPEA shift of zero, an imaginary level shift of 0.3~a.u., and the Douglas-Kroll-Hess scalar-relativistic Hamiltonian.

\footnotesize

\bigskip
\begin{minipage}{0.45\textwidth}
\begin{verbatim}
12
S0 min -3017.895875
C +2.593512 -0.222280 +0.038020
N +1.817239 -1.374052 +0.106087
N +1.820018 +0.928609 +0.016124
C +0.420515 +1.060202 +0.036590
C -0.276129 -0.221097 +0.104128
C +0.445075 -1.374668 +0.137672
O +3.809942 -0.230391 +0.000890
O -0.095280 +2.172358 -0.005543
H +2.341064 +1.798650 -0.039884
H -0.025820 -2.348442 +0.187033
H +2.332546 -2.243385 +0.126974
Br -2.161345 -0.225354 +0.151685
\end{verbatim}
\end{minipage}

\bigskip
\begin{minipage}{0.45\textwidth}
\begin{verbatim}
12
S1 min -3017.744217
C +2.577804 -0.220493 +0.048783
N +1.822291 -1.366540 +0.125086
N +1.825424 +0.953874 +0.071580
C +0.432186 +0.959745 +0.047299
C -0.273703 -0.210910 +0.098812
C +0.423043 -1.426108 +0.144118
O +3.800354 -0.208315 -0.019216
O -0.093784 +2.211681 +0.007893
H +2.344346 +1.806094 -0.093610
H -0.040283 -2.398408 +0.190024
H +2.364984 -2.218696 +0.114489
Br -2.161324 -0.161773 +0.124516
\end{verbatim}
\end{minipage}

\bigskip
\begin{minipage}{0.45\textwidth}
\begin{verbatim}
12
T1 min -3017.786734
C +2.499331 -0.231593 -0.099666
N +1.732554 -1.384696 -0.033702
N +1.812776 +0.923500 +0.285298
C +0.421713 +1.095003 +0.189546
C -0.293909 -0.176691 +0.248591
C +0.490247 -1.379361 +0.580637
O +3.669919 -0.223411 -0.447543
O -0.075167 +2.200676 +0.072047
H +2.357644 +1.776040 +0.198077
H +0.041483 -2.325239 +0.845078
H +2.240934 -2.238646 -0.228181
Br -1.876188 -0.315430 -0.750410
\end{verbatim}
\end{minipage}

\bigskip
\begin{minipage}{0.45\textwidth}
\begin{verbatim}
12
S2S1 CoIn -3017.740644 & -3017.740562
C +2.602825 -0.130131 -0.010764
N +1.851709 -1.296877 +0.057084
N +1.874572 +1.013291 +0.323128
C +0.461426 +1.059319 +0.151578
C -0.230134 -0.139362 +0.238188
C +0.491333 -1.359579 +0.340157
O +3.794219 -0.138639 -0.287726
O -0.034278 +2.247471 -0.078423
H +2.364169 +1.880308 +0.128134
H +0.049789 -2.327331 +0.520621
H +2.384101 -2.143333 -0.118367
Br -2.104540 -0.178272 +0.099673
\end{verbatim}
\end{minipage}

\bigskip
\begin{minipage}{0.45\textwidth}
\begin{verbatim}
12
S1S0 CoIn -3017.743518 & -3017.743082
C +1.958439 -0.237647 -0.312087
N +1.277050 -1.355149 +0.318549
N +1.567098 +0.977143 +0.166224
C +0.208320 +1.317503 +0.548945
C -0.580797 +0.130594 +0.975619
C +0.358292 -1.073708 +1.209022
O +2.867209 -0.470314 -1.077274
O -0.088772 +2.472018 +0.495973
H +2.084834 +1.758582 -0.221988
H +0.224230 -1.759596 +2.043254
H +1.715915 -2.261891 +0.157601
Br -1.400332 -0.298246 -0.731076
\end{verbatim}
\end{minipage}

\bigskip
\begin{minipage}{0.45\textwidth}
\begin{verbatim}
12
S1T2 MXP -3017.739802 & -3017.740501
C +2.595622 -0.214758 +0.045562
N +1.803785 -1.362209 +0.106812
N +1.807845 +0.943051 +0.006083
C +0.437529 +0.926977 +0.041314
C -0.274910 -0.202139 +0.102601
C +0.437492 -1.429393 +0.130401
O +3.831481 -0.219290 +0.017289
O -0.071971 +2.236513 -0.004470
H +2.302539 +1.825280 -0.055395
H -0.029713 -2.401319 +0.186206
H +2.344053 -2.220627 +0.125332
Br -2.162415 -0.161934 +0.158037
\end{verbatim}
\end{minipage}

\bigskip
\begin{minipage}{0.45\textwidth}
\begin{verbatim}
12
T2T1 CoIn -3017.744857 & -3017.744344
C +2.568862 -0.214887 -0.014905
N +1.815637 -1.362883 -0.028394
N +1.819151 +0.966966 +0.018139
C +0.423228 +0.988835 +0.107446
C -0.273986 -0.222422 +0.154128
C +0.423104 -1.398069 +0.160784
O +3.791323 -0.198527 -0.015069
O -0.050423 +2.138831 -0.464294
H +2.344595 +1.773021 +0.333976
H -0.024265 -2.371231 +0.303216
H +2.352454 -2.217412 -0.003093
Br -2.169208 -0.187459 +0.188062
\end{verbatim}
\end{minipage}

\bigskip
\begin{minipage}{0.45\textwidth}
\begin{verbatim}
12
S1 min (CBr=2.8A,CCBr=119deg) -3017.759514
C +2.575721 -0.227511 +0.048165
N +1.817604 -1.387343 +0.112008
N +1.817212 +0.929795 -0.019866
C +0.411261 +1.070983 -0.033571
C -0.238859 -0.222096 +0.039953
C +0.429163 -1.384103 +0.109457
O +3.797684 -0.232333 +0.052002
O -0.108231 +2.174423 -0.099108
H +2.353097 +1.791062 -0.068135
H -0.047182 -2.357597 +0.165717
H +2.344171 -2.248447 +0.164055
Br -3.035882 -0.333468 +0.105214
\end{verbatim}
\end{minipage}



\clearpage
\section*{ADC(2) Geometries}

The following geometries of 5-bromouracil were optimized with ADC(2)/def2-TZVP.
The calculations were performed with \textsc{Turbomole} 7.0.

\footnotesize

\bigskip
\begin{minipage}{0.45\textwidth}
\begin{verbatim}
12
S0 min ADC E -2986.073995
C +2.580674 -0.222147 +0.036374
N +1.814023 -1.370339 +0.104824
N +1.817193 +0.927940 +0.005355
C +0.424728 +1.063791 +0.033616
C -0.265098 -0.223943 +0.106217
C +0.444288 -1.372420 +0.138493
O +3.794622 -0.231873 +0.007035
O -0.100238 +2.158813 -0.000869
H +2.342075 +1.794628 -0.044994
H -0.031291 -2.343330 +0.192140
H +2.332558 -2.236405 +0.129566
Br -2.132197 -0.224563 +0.152016
\end{verbatim}
\end{minipage}

\bigskip
\begin{minipage}{0.45\textwidth}
\begin{verbatim}
12
S1 min -2985.947347
C +2.572216 -0.226718 +0.038338
N +1.821375 -1.366984 +0.116103
N +1.809927 +0.943550 +0.027503
C +0.432516 +0.923844 +0.043669
C -0.270760 -0.207235 +0.118585
C +0.432127 -1.444326 +0.203335
O +3.789193 -0.197498 -0.014156
O -0.083544 +2.232971 -0.010367
H +2.320046 +1.810850 -0.076092
H -0.032754 -2.414916 +0.111590
H +2.373857 -2.213003 +0.151728
Br -2.142862 -0.120382 +0.149536
\end{verbatim}
\end{minipage}

\bigskip
\begin{minipage}{0.45\textwidth}
\begin{verbatim}
12
T1 boat min E -2985.966263
C +2.517907 -0.222616 -0.062806
N +1.743334 -1.374048 -0.037749
N +1.816972 +0.929301 +0.263594
C +0.424199 +1.088814 +0.135863
C -0.275374 -0.170740 +0.223166
C +0.494635 -1.396974 +0.559840
O +3.702076 -0.234652 -0.336554
O -0.076090 +2.191878 -0.030395
H +2.357507 +1.784186 +0.186052
H +0.006360 -2.354462 +0.668371
H +2.269612 -2.224523 -0.199176
Br -1.959803 -0.296013 -0.510433
\end{verbatim}
\end{minipage}

\bigskip
\begin{minipage}{0.45\textwidth}
\begin{verbatim}
12
S2S1 CoIn -2985.921551 & -2985.921540
C +2.534856 -0.128550 -0.115879
N +1.752915 -1.308688 -0.092509
N +1.856915 +0.963640 +0.379559
C +0.444098 +1.068760 +0.252863
C -0.245267 -0.074812 +0.302333
C +0.542989 -1.362028 +0.493095
O +3.692705 -0.143471 -0.482678
O +0.006374 +2.282223 +0.006815
H +2.364813 +1.840400 +0.328151
H +0.212766 -2.245928 +1.026727
H +2.292815 -2.147233 -0.295139
Br -1.950788 -0.257452 -0.440057
\end{verbatim}
\end{minipage}

\bigskip
\begin{minipage}{0.45\textwidth}
\begin{verbatim}
12
S0S1 CoIn -2985.945232 & -2985.945075
C +2.040450 -0.247992 -0.279204
N +1.307104 -1.332946 +0.260589
N +1.576135 +0.994624 +0.105791
C +0.257923 +1.308362 +0.525088
C -0.496458 +0.103825 +0.893587
C +0.324405 -1.102057 +1.148600
O +3.029407 -0.434097 -0.952770
O -0.142133 +2.450119 +0.472938
H +2.121877 +1.772710 -0.252260
H +0.160186 -1.794935 +1.968451
H +1.779902 -2.227570 +0.165520
Br -1.767312 -0.290752 -0.483568
\end{verbatim}
\end{minipage}

\bigskip
\begin{minipage}{0.45\textwidth}
\begin{verbatim}
12
S1T2 MXP -2985.944987 & -2985.944989
C +2.570313 -0.223733 +0.041825
N +1.816065 -1.368019 +0.122823
N +1.814081 +0.942101 +0.025185
C +0.430302 +0.958256 +0.043548
C -0.268392 -0.189452 +0.134040
C +0.434873 -1.448717 +0.237373
O +3.787353 -0.212551 -0.011696
O -0.089620 +2.213654 -0.023108
H +2.323303 +1.808918 -0.087658
H -0.037603 -2.402791 +0.049902
H +2.371959 -2.212662 +0.157811
Br -2.131296 -0.144852 +0.169728
\end{verbatim}
\end{minipage}

\bigskip
\begin{minipage}{0.45\textwidth}
\begin{verbatim}
12
T2T1 CoIn -2985.946820 & -2985.946713
C +2.571396 -0.222908 +0.046384
N +1.817983 -1.368897 +0.124005
N +1.815480 +0.938099 -0.042528
C +0.428195 +0.967882 +0.019720
C -0.269654 -0.192639 +0.089747
C +0.432034 -1.446513 +0.129387
O +3.789550 -0.211470 +0.047727
O -0.095210 +2.208029 -0.011072
H +2.329937 +1.807479 -0.092495
H -0.040664 -2.414863 +0.093375
H +2.371379 -2.213114 +0.185408
Br -2.129954 -0.156321 +0.150339
\end{verbatim}
\end{minipage}

\bigskip
\begin{minipage}{0.45\textwidth}
\begin{verbatim}
12
S1 min (CBr=2.8A,CCBr=119deg) -2985.923206
C +2.561966 -0.232482 +0.048026
N +1.800741 -1.390520 +0.110880
N +1.813216 +0.934712 -0.023850
C +0.411727 +1.109726 -0.035023
C -0.189757 -0.202767 +0.038700
C +0.423150 -1.384774 +0.110211
O +3.776457 -0.243798 +0.054817
O -0.104855 +2.203596 -0.097974
H +2.366988 +1.785882 -0.069443
H -0.089616 -2.338709 +0.166642
H +2.326591 -2.252538 +0.165033
Br -2.980848 -0.414961 +0.107873
\end{verbatim}
\end{minipage}


\clearpage
\section*{MRCIS Geometries}

The following geometries of 5-bromouracil were optimized with(10,8)/cc-pVDZ-DK.
The calculations were performed with \textsc{Columbus} 7.0.
The orbitals came from a SA-CASSCF(10,8)/cc-pVDZ-DK calculation, averaging over four singlets and three triplets simultaneously.
For the MRCI, the CAS(10,8) reference space was restricted to have at most two electrons in the three antibonding orbitals.
22 inner core orbitals were kept frozen in the MRCI step.

\footnotesize

\bigskip
\begin{minipage}{0.45\textwidth}
\begin{verbatim}
12
S0 min -3016.549508
C +2.574450 -0.219017 +0.036389
N +1.825363 -1.365782 +0.104347
N +1.828072 +0.929069 +0.005037
C +0.432268 +1.060167 +0.033578
C -0.264690 -0.224419 +0.106225
C +0.445286 -1.374716 +0.138568
O +3.766419 -0.227452 +0.007540
O -0.086893 +2.144120 -0.000555
H +2.351888 +1.781537 -0.044666
H -0.025482 -2.345242 +0.192082
H +2.341601 -2.219713 +0.128623
Br -2.166946 -0.218400 +0.152605
\end{verbatim}
\end{minipage}

\bigskip
\begin{minipage}{0.45\textwidth}
\begin{verbatim}
12
S1 min -3016.395668
C +2.569456 -0.225511 +0.035906
N +1.838669 -1.368012 +0.092070
N +1.822056 +0.938395 +0.011582
C +0.429273 +0.959343 +0.036300
C -0.269690 -0.210921 +0.090434
C +0.431750 -1.431660 +0.118589
O +3.762330 -0.195535 +0.008868
O -0.072698 +2.186075 +0.002519
H +2.342432 +1.789828 -0.036626
H -0.032394 -2.399892 +0.166972
H +2.378557 -2.207030 +0.109257
Br -2.178674 -0.173655 +0.125850
\end{verbatim}
\end{minipage}

\bigskip
\begin{minipage}{0.45\textwidth}
\begin{verbatim}
12
T1 min -3016.434611
C +2.533945 -0.225142 +0.035106
N +1.778337 -1.381540 +0.093391
N +1.801300 +0.914151 -0.223077
C +0.431107 +1.047647 +0.096186
C -0.286127 -0.203179 -0.000197
C +0.493219 -1.399435 -0.425470
O +3.713832 -0.225649 +0.183786
O -0.039142 +2.122535 +0.402020
H +2.334410 +1.760420 -0.157414
H +0.023502 -2.354227 -0.594668
H +2.316022 -2.221258 +0.176555
Br -1.889308 -0.408809 +0.963251
\end{verbatim}
\end{minipage}

\bigskip
\begin{minipage}{0.45\textwidth}
\begin{verbatim}
12
S2S1 CoIn -3016.365491 & -3016.365381
C +2.418778 -0.150099 +0.175104
N +1.613358 -1.290422 +0.155870
N +1.805746 +0.955506 -0.359270
C +0.397626 +1.114095 -0.246975
C -0.345117 -0.045038 -0.438495
C +0.456572 -1.300023 -0.548679
O +3.535827 -0.184035 +0.578735
O -0.015384 +2.266854 +0.034734
H +2.347690 +1.795356 -0.284497
H +0.032114 -2.246732 -0.846848
H +2.068746 -2.128093 +0.468190
Br -2.002719 -0.263886 +0.539666
\end{verbatim}
\end{minipage}

\bigskip
\begin{minipage}{0.45\textwidth}
\begin{verbatim}
12
S1T2 MXP -3016.392558 & -3016.392484
C +2.542485 -0.235116 +0.010844
N +1.807318 -1.365588 +0.078632
N +1.802180 +0.950274 -0.064450
C +0.427783 +0.948053 +0.150295
C -0.271978 -0.203843 +0.335231
C +0.406145 -1.398965 +0.316927
O +3.734190 -0.194134 -0.000419
O -0.031678 +2.214691 +0.073622
H +2.339464 +1.760556 +0.178062
H -0.050863 -2.364610 +0.442677
H +2.344631 -2.199227 +0.188270
Br -2.179136 -0.087942 +0.565407
\end{verbatim}
\end{minipage}

\bigskip
\begin{minipage}{0.45\textwidth}
\begin{verbatim}
12
T2T1 CoIn -3016.401632 & -3016.401202
C +2.546912 -0.231906 +0.024953
N +1.805390 -1.366340 +0.096412
N +1.815148 +0.945626 -0.058553
C +0.434141 +0.974150 +0.161032
C -0.271373 -0.190532 +0.326630
C +0.405801 -1.398699 +0.302479
O +3.739788 -0.212503 +0.012938
O -0.072236 +2.205454 +0.119208
H +2.357057 +1.769495 +0.110923
H -0.060963 -2.360451 +0.419981
H +2.340130 -2.202139 +0.202113
Br -2.169255 -0.108008 +0.556983
\end{verbatim}
\end{minipage}



\clearpage
\section{Other Geometries}

This geometry was reported by Kobylecka et al.\cite{Kobylecka2009JPCA} as the $S_0$ minimum of 5-bromouracil.
It was computed at the CASSCF(12,10)/6-311G* level of theory, using a non-relativistic Hamiltonian.

\footnotesize

\bigskip
\begin{minipage}{0.45\textwidth}
\begin{verbatim}
12
Kobylecka CASSCF(12,10)/6-311G*
C +0.000000 +0.000000 +0.000000
N +0.000000 +0.000000 +1.393026
N +1.280253 +0.000000 -0.530377
C +2.520387 +0.000000 +0.139797
C +2.367687 +0.000000 +1.601476
C +1.140042 +0.000000 +2.158851
O -1.012793 +0.000000 -0.659778 
O +3.558429 +0.000000 -0.477993 
H +1.331866 +0.000000 -1.540858 
H +0.988397 +0.000000 +3.230359 
H -0.911372 +0.000000 +1.824706 
Br +3.934298 +0.000000 +2.662710
\end{verbatim}
\end{minipage}


\bigskip
\noindent
This geometry was optimized at the MP2/ANO-RCC-VDZP level of theory, using \textsc{Gaussian} 09.
The corresponding frequency calculation was employed for the generation of the initial conditions for the dynamics simulations.

\footnotesize

\bigskip
\begin{minipage}{0.45\textwidth}
\begin{verbatim}
12
Kobylecka CASSCF(12,10)/6-311G*
C +0.000000 +0.000000 +0.000000
N +0.000000 +0.000000 +1.393026
N +1.280253 +0.000000 -0.530377
C +2.520387 +0.000000 +0.139797
C +2.367687 +0.000000 +1.601476
C +1.140042 +0.000000 +2.158851
O -1.012793 +0.000000 -0.659778 
O +3.558429 +0.000000 -0.477993 
H +1.331866 +0.000000 -1.540858 
H +0.988397 +0.000000 +3.230359 
H -0.911372 +0.000000 +1.824706 
Br +3.934298 +0.000000 +2.662710
\end{verbatim}
\end{minipage}


%
%

\end{document}